\documentclass[doublecol]{epl2}

\usepackage{graphicx}
\usepackage{amsmath}
\usepackage{color}
\usepackage{hyperref}

\title{Closing the pseudogap quietly.}
\shorttitle{Closing the pseudogap quietly.} 

\author{J.G. Storey\inst{1}}
\shortauthor{J.G. Storey}

\institute{                    
  \inst{1} Robinson Research Institute, Victoria University - 
P.O. Box 600, Wellington, New Zealand
}


\abstract{
The physical properties of hole-doped cuprate high-temperature superconductors 
are heavily influenced by an energy gap known as the pseudogap whose origin 
remains a mystery second only to that of superconductivity itself. A key 
question is whether the pseudogap closes at a temperature $T^*$. The 
absence of a specific heat anomaly, together with persistent entropy losses up 
to 300K, have long suggested that the pseudogap does not vanish at $T^*$. 
However, amid a growing body of evidence from other techniques pointing to the 
contrary we revisit this question. Here we investigate if, by adding a 
temperature dependence to the pseudogap energy and quasiparticle lifetime in 
the resonating-valence-bond spin-liquid model of Yang Rice and Zhang, we can 
close the pseudogap quietly in the specific heat. 
}

\begin{document}

\maketitle

The physical properties of hole-doped cuprate high-temperature superconductors are strongly influenced, over a wide range of temperature and doping, by a depletion in the electronic density of states known as the pseudogap. In momentum-space it is manifest as a gapping of the large hole-like Fermi surface near the antinodal regions of the Brillouin zone, at ($\pm\pi$,0) and (0,$\pm\pi$), leaving behind ungapped ``Fermi arcs''\cite{FERMIARCS}. The origin of the pseudogap remains a mystery second only to that of high-temperature superconductivity itself, and it is widely hoped that by investigating the former we might uncover valuable insights for understanding the latter. A key question is whether the pseudogap closes at a temperature $T^*$. In recent years, evidence has been building that suggests that it does. These include abrupt changes in the Kerr Effect\cite{XIA}, time resolved reflectivity\cite{HE}, as well as the direct observation of a reconstruction of the antinodal electronic structure by angle-resolved photoemission spectroscopy\cite{HASHIMOTO1,HE} (ARPES).
In this work, we aim to reconcile those results with thermodynamic measurements, in particular the electronic entropy and specific heat, which have long suggested that the pseudogap is temperature independent\cite{LORAM,ENTROPYLSCO,LORAMPG,ENTROPYDATA2}.

By way of introduction, the electronic entropy is defined as $S(T)=-2k_B\int{f_w(E,T)N(E)dE}$\cite{ENTROPY} where $N(E)$ is the density of states, and $f_w(E,T)$ is a ``Fermi window'' which expands with temperature and is related to the Fermi distribution function $f$ by $f\ln f+(1-f)ln(1-f)$. Put simply, $S(T)$ is a count of the thermally active states. The electronic specific heat coefficient is given by the temperature derivative of the entropy, $\gamma(T)=\partial S(T)/\partial T$.
Three apparently universal observations have been made from high-resolution differential specific heat studies on a variety of hole-doped cuprates\cite{LORAM93,LORAM,ENTROPYLSCO,LORAMPG,ENTROPYDATA2}. These are:
i) a loss of entropy in low to slightly overdoped samples, that persists right up to the highest temperatures measured. The entropy decreases at a rate of about 1 $k_B$ per doped hole;
ii) a collapse in the magnitude of the specific heat jump, $\Delta\gamma$, at $T_c$ below a critical doping of 0.19 holes/Cu;
 and iii) a smooth downturn in the normal-state electronic specific heat with no specific heat jump at $T^*$.
These features were originally modeled by Loram in terms of a temperature-independent non-states-conserving V-shaped gap, pinned to the Fermi level ($E_F$) of a flat density of states. The gap widens with reducing doping\cite{ENTROPYDATA2}. In contrast to the superconducting gap, where the low-energy states are pushed just above the gap edge, it is surmised in this model that the pseudogap redistributes those states to much higher energies. In this scenario $T^*$ represents an energy scale where thermal fluctuations become comparable in magnitude to the size of the pseudogap, rather than a phase transition temperature. If one tries to fill in such a pseudogap with temperature, thereby simulating expanding Fermi arcs\cite{FERMIARCS,FERMIARCS2}, problems arise. Firstly, the lost entropy is eventually recovered, contradicting (i). Secondly, a kink in the entropy appears at $T^*$ together with a corresponding jump in the heat capacity\cite{STOREYRAMAN}, contradicting (iii). And finally, we might expect a double-peak structure to appear in the superconducting anomaly near critical doping, where $T^*$ is less than $T_c$, altering the doping dependence of $\Delta\gamma(T_c)$ compared to (ii). But perhaps this just means that this model is incomplete, and if so, what are we missing?

\begin{figure*}
\centering
\includegraphics[width=\linewidth]{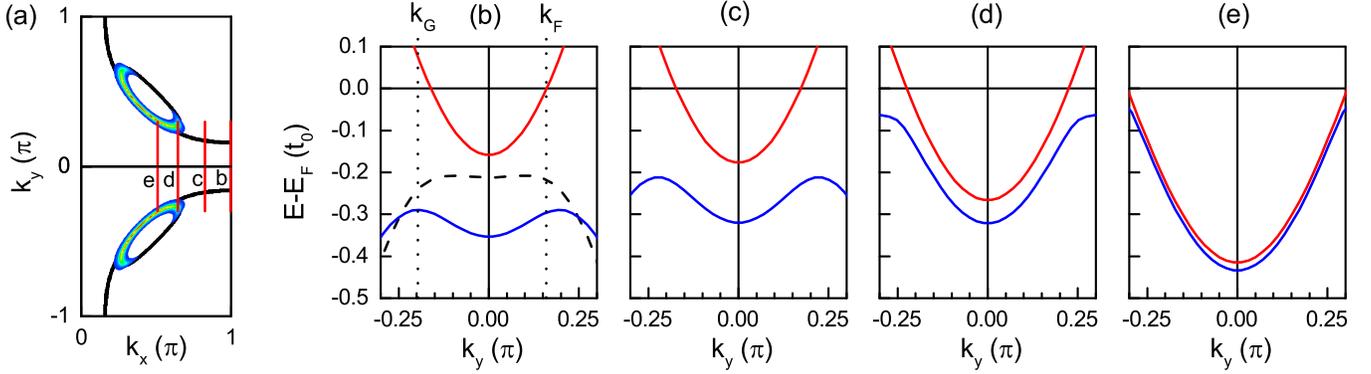}%
\caption{
(Color online) (a) Spectral weight at the Fermi level calculated from the YRZ model for $x$=0.12. The black curves indicate the back of the nodal hole pocket, as well as the Fermi level crossings when $E_g$=0.  Dispersions calculated along the vertical momentum-space cuts are shown in plots (b) to (e), both with (blue line) and without (red line) the pseudogap. In the $k_x=\pi$ cut, (b), the dashed line shows the Bogoliubov dispersion from the upper YRZ band ($\Delta_0$=0.12$t_0$). These results closely reproduce the experimental data of Refs.~\cite{HASHIMOTO1} and \cite{HE}
} 
\label{EKFIG}
\end{figure*}

In the following we will investigate the effects of a tight-binding density of states, thermal lifetime broadening, and the combination of these with a Fermi-surface reconstruction model for the pseudogap given by the resonating valence bond spin liquid ansatz of Yang, Rice and Zhang (YRZ)\cite{YRZ}.
Detailed descriptions of the YRZ model have been published several times\cite{YRZ,SCHACHINGER,BORNE1}, but for completeness we briefly list the equations used in this work.
In the normal state the coherent part of the electron Green's function is given by
\begin{equation}
G(\textbf{k},\omega,x)=\frac{g_t(x)}{\omega-\xi_\textbf{k}-\frac{E_g^2(\textbf{k})}{\omega+\xi_\textbf{k}^0}}
\label{eq:GYRZ}
\end{equation}
where $\xi_\textbf{k}=-2t(x)(\cos k_x+\cos k_y)
-4t^\prime(x)\cos k_x\cos k_y
-2t^{\prime\prime}(x)(\cos 2k_x+\cos 2k_y)-\mu_p(x)$ is the tight-binding energy-momentum dispersion, $\xi_\textbf{k}^0=-2t(x)(\cos k_x+\cos k_y )$ is the nearest-neighbour term, and $E_g(\textbf{k})=[E_g^0(x)/2](\cos k_x-\cos k_y)$
is the pseudogap. The chemical potential $\mu_p(x)$ is chosen according to the Luttinger sum rule.
The doping-dependent coefficients are given by $t(x)=g_t(x)t_0+(3/8)g_s(x)J\chi$, $t^\prime(x)=g_t(x)t_0^\prime$ and $t^{\prime\prime}(x)=g_t(x)t_0^{\prime\prime}$, where $g_t(x)=2x/(1+x)$ and $g_s(x)=4/(1+x)^2$ are the Gutzwiller factors. The bare parameters $t^\prime/t_0=-0.3$, $t^{\prime\prime}/t_0=0.2$, $J/t_0=1/3$ and $\chi=0.338$ are the same as used previously\cite{YRZ}.
Equation~\ref{eq:GYRZ} can be re-written as
\begin{equation}
G(\textbf{k},\omega,x)=\sum_{\alpha=\pm}{\frac{g_t(x)W_\textbf{k}^\alpha(x)}{\omega-E_\textbf{k}^\alpha(x)}}
\label{eq:GYRZ2}
\end{equation}
where the energy-momentum dispersion is reconstructed by the pseudogap into upper and lower branches
\begin{equation}
E_\textbf{k}^\pm=\frac{1}{2}(\xi_\textbf{k}-\xi_\textbf{k}^0)\pm\sqrt{\left(\frac{\xi_\textbf{k}+\xi_\textbf{k}^0}{2}\right)^2+E_g^2(\textbf{k})}
\label{eq:EK}
\end{equation}
that are weighted by
\begin{equation}
W_\textbf{k}^\pm=\frac{1}{2}\left[1\pm\frac{(\xi_\textbf{k}+\xi_\textbf{k}^0)/2}{\sqrt{[(\xi_\textbf{k}+\xi_\textbf{k}^0)/2]^2+E_g^2(\textbf{k})}}\right]
\label{eq:WK}
\end{equation}
In the superconducting state there are four energy branches $\pm E_S^\alpha=\pm\sqrt{(E_\textbf{k}^\alpha)^2+\Delta_\textbf{k}^2}$, where $\alpha=\pm$ and $\Delta_\textbf{k}=[\Delta_0(x)/2](\cos k_x-\cos k_y)$ is the superconducting gap.
The density of states (DOS), from which the entropy and heat capacity can be calculated, is given by
\begin{equation}
N(\omega)=\sum_{\alpha=\pm,\textbf{k}}{g_t(x)W_\textbf{k}^\alpha[(u_\textbf{k}^\alpha)^2\delta(\omega-E_S^\alpha)+(v_\textbf{k}^\alpha)^2\delta(\omega+E_S^\alpha)]}
\label{eq:DOS}
\end{equation}
where $(u_\textbf{k}^\alpha)^2=0.5(1+E_\textbf{k}^\alpha/E_S^\alpha)$ and $(v_\textbf{k}^\alpha)^2=0.5(1-E_\textbf{k}^\alpha/E_S^\alpha)$ are the Bogoliubov weights.

The reason for choosing this model is because it successfully describes experimental data from a wide range of techniques\cite{RICE}, including the specific heat\cite{LEBLANC1,BORNE2}. However, the previous works did not consider a temperature dependent pseudogap term. In fig.~\ref{EKFIG} we plot the calculated energy momentum dispersion in the superconducting state along cuts in the $k_y$ direction near the antinodes for $x$=0.12, both with, and without ($E_g$=0) the pseudogap. The results reproduce the ARPES-derived dispersions measured below and above $T^*$ respectively\cite{HASHIMOTO1,HE}, providing compelling evidence for the closure of the pseudogap at $T^*$. Key details are reproduced such as the separation between the momentum of the minimum binding energy of the dispersion $k_G$ from the Fermi momentum $k_F$, a signature of non-particle-hole symmetric order\cite{HASHIMOTO1}. Moreover we can identify the flat dispersion of the shoulder feature observed in ARPES energy dispersion curves\cite{HE} as belonging to the Bogoliubov dispersion arising from the upper YRZ band, $-\sqrt{(E_\textbf{k}^+)^2+\Delta_\textbf{k}^2}$.

\begin{figure*}
\centering
\includegraphics[width=\linewidth]{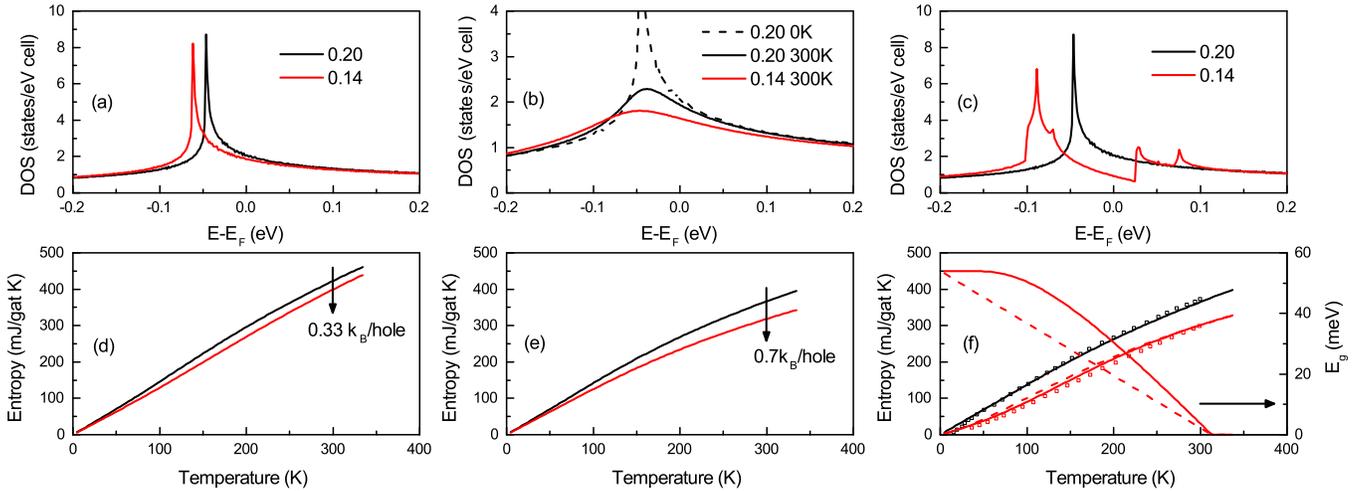}%
\caption{
(Color online) Density of states for $x$=0.20 and 0.14 in the case of: (a) a rigid shift of the Fermi level and no lifetime broadening; (b) the addition of thermal lifetime broadening terms $\pi k_B T$ and $2\pi k_B T$ respectively; and (c) a YRZ-like reconstruction. The corresponding electronic entropies are shown in plots (d) to (f). Plot (f) includes experimental data for La$_{2-x}$Sr$_x$CuO$_4$\cite{ENTROPYLSCO} and the calculated curves include the thermal broadening terms. The two fits to the $x$=0.14 data  correspond to the two pseudogap temperature dependences of $E_g(T)$.
} 
\label{ENTROPYFIG}
\end{figure*}

Since we wish to understand the effect of adding a temperature dependence to $E_g$, from here onwards we fix the tight binding coefficients to their values at $x$=0.20 and neglect the $g_t(x)$ prefactor in the equation for the density of states. (Normally the $x$ dependence of these terms, which narrow the bands but reduce the magnitude of the DOS, would complement rather than counteract the pseudogap.) To fit the experimental entropy data $t_0$ is set to 0.225 eV. Beginning for a moment without the pseudogap, the defining feature of the tight-binding DOS is the van Hove singularity (vHs) located just below $E_F$ for $x$=0.20 (see fig.~\ref{ENTROPYFIG}(a)). Assuming a rigid shift of $E_F$ away from the vHs with decreasing doping results in a persistent decrease in entropy, as shown in fig.~\ref{ENTROPYFIG}(d). However at 300 K, the rate of decrease is only 0.33 $k_B$/hole compared with the observed 1 $k_B$/hole\cite{ENTROPYDATA2}.

Lifetime broadening can also affect the high-temperature heat capacity, and hence the entropy, by smoothing features in the DOS\cite{THIESSEN}. From resistivity measurements\cite{HUSSEY4} we infer a linear-in-temperature scattering rate (inverse lifetime) given by $\Gamma = 0.01t_0+\beta k_B T$, with a slope $\beta$ that increases with decreasing doping. The most computationally efficient way of incorporating this term is by convolving the DOS with the lorentzian $\Gamma/\pi[(\omega-E)^2+\Gamma^2]$. Figure~\ref{ENTROPYFIG}(b) illustrates the thermally broadened vHs at 300 K for $x$=0.20 and 0.14 with $\beta$=1 and 2 respectively. The entropy decrease is now larger at 0.7 $k_B$/hole (fig.~\ref{ENTROPYFIG}(e)), but it is still not enough, especially at low temperatures. This necessitates the incorporation of a pseudogap.

In fig.~\ref{ENTROPYFIG}(c) we add a pseudogap for $x$=0.14 by setting $E_g^0$=54 meV. Based on the ARPES results we initially assume that the pseudogap closes linearly with temperature according to $E_g(T)=E_g^0-2k_BT$. The van Hove singularity and lifetime broadening effects are also included. The calculated entropy compares well with experimental data for La$_{2-x}$Sr$_x$CuO$_4$\cite{ENTROPYLSCO}, shown in fig.~\ref{ENTROPYFIG}(f), with the entropy decrease approaching the observed 1 $k_B$/hole. The low-temperature fit can be further improved by taking a more gradual initial $T$-dependence given by
\begin{equation}
E_g(T) = E_g^0\left[2-1/\tanh\left(\frac{E_g^0\ln 3}{4k_B T}\right)\right]
\label{eq:Egtanh}
\end{equation}

\begin{figure}
\centering
\includegraphics[width=\linewidth]{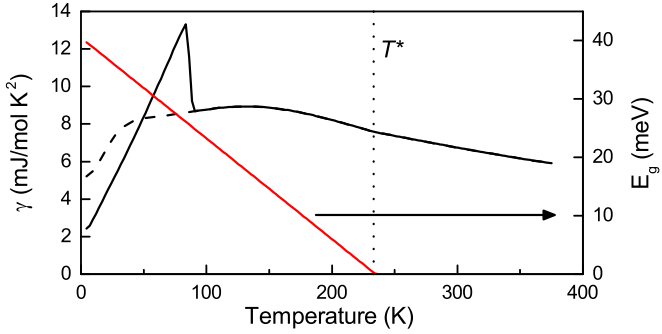}%
\caption{
(Color online) Electronic specific heat for a YRZ-like pseudogap that closes as $E_g=E_g^0-2k_B T$ in the presence of lifetime broadening $\Gamma = 0.01t_0+k_B T$. Note the absence of a specific heat jump at $T^*$.
} 
\label{GAMMAFIG}
\end{figure}

\begin{figure}
\centering
\includegraphics[width=\linewidth]{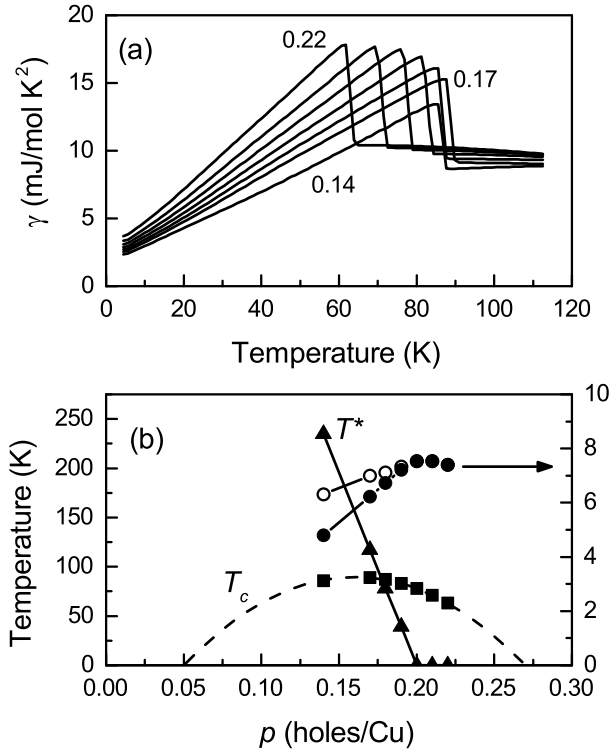}%
\caption{
(a) Superconducting electronic specific heat jump for $x$=0.14, and 0.17 to 0.22, in the presence of a YRZ-like pseudogap that closes as $E_g=E_g^0-2k_B T$ and lifetime broadening $\Gamma = 0.01t_0+k_B T$. (b) Doping dependence of $T_c$ (squares), $T^*$ (triangles), and the specific heat jump at $T_c$ both with (filled circles) and without (empty circles) the pseudogap.
} 
\label{DGAMMAFIG}
\end{figure}

We now turn to the specific heat coefficient, $\gamma$. Figure~\ref{GAMMAFIG} shows $\gamma(T)$ calculated for a 40meV pseudogap which closes linearly with temperature in the presence of lifetime broadening with $\beta$=1. (This of course does not contain the additional experimentally observed contributions from fluctuations near $T_c$.) There is no obvious jump at $T^*$, only a slight change in slope. If $E_g(T)$ was rounded near $T^*$, due to doping inhomogeneity for example, the specific heat would become even smoother there. Finally, in fig.~\ref{DGAMMAFIG} we plot the doping dependence of the specific heat jump at $T_c$ assuming a parabolic superconducting gap doping dependence $\Delta(x)=0.103t_0[1-82.6(x-0.16)^2]$, and the YRZ pseudogap doping dependence $E_g^0(x)=3t_0(0.2-x)$ for $x\leq 0.2$. Here we take the closure of the pseudogap to lie at $x=0.2$ in continuity with YRZ, however it has been extensively shown that this occurs at slightly lower doping $x=0.19$\cite{OURWORK1}. The pseudogap model reproduces the collapse of the specific heat jump as reported for example in refs.~\cite{LORAM93} and \cite{ENTROPYDATA2}. Note that here we have taken a doping independent lifetime broadening, $\beta=1$. Increasing $\beta$ with decreasing doping would increase the rate of collapse of $\Delta\gamma(T_c)$.

To conclude, the absence of a specific heat jump at $T^*$, together with persistent losses in entropy at high temperatures, has long been taken as evidence that the pseudogap does not close there. Driven by a growing body of evidence from other experimental probes pointing to the contrary we have explored this question. By adding a linear-in-temperature scattering rate to a YRZ-like reconstruction model, it is possible to close the pseudogap quietly in the specific heat. A similar result is expected for the antiferromagnetic Brillouin-zone-folding Fermi surface reconstruction model\cite{CHUBUKOV}. The entropy recovery expected from the closing gap is offset by scattering-induced broadening of the van Hove singularity. This scenario could be tested experimentally by searching for an ongoing divergence between neighbouring entropy curves above $T^*$.

\begin{acknowledgments}
Supported by the Marsden Fund Council from Government funding, administered by the Royal Society of New Zealand.
\end{acknowledgments}

\bibliographystyle{eplbib}

\end{document}